\documentclass[twocolumn]{aastex7}

\usepackage{amsmath}

\def\muG{{\mu\rm G}}

\received{}
\revised{}
\accepted{}

\submitjournal{The Astrophysical Journal}

\shorttitle{Ultra-High-Energy Cosmic Rays from Nearby Radio Galaxies}
\shortauthors{Seo et al.}

\begin{document}

\title{Energy Spectrum and Mass Composition of Ultra-High-Energy Cosmic Rays Originating from Relativistic Jets of Nearby Radio Galaxies}

\author[0000-0002-5550-8667]{Jeongbhin Seo}
\affiliation{Department of Physics, College of Natural Sciences, UNIST, Ulsan 44919, Korea}
\affiliation{Los Alamos National Laboratory, Theoretical Division, Los Alamos, NM 87545, USA}
\email{jseo@lanl.gov}

\author[0000-0002-5455-2957]{Dongsu Ryu}
\affiliation{Department of Physics, College of Natural Sciences, UNIST, Ulsan 44919, Korea}
\email{dsryu@unist.ac.kr}

\author[0000-0002-4674-5687]{Hyesung Kang}
\affiliation{Department of Earth Sciences, Pusan National University, Busan 46241, Korea}
\email{hskang@pusan.ac.kr}

\correspondingauthor{Dongsu Ryu}\email{dsryu@unist.ac.kr}
\correspondingauthor{Hyesung Kang}\email{hskang@pusan.ac.kr}

\begin{abstract}


Relativistic jets of radio galaxies (RGs) are possible sources of ultra-high-energy cosmic rays (UHECRs). Recent studies combining relativistic hydrodynamic simulations with Monte Carlo particle transport have demonstrated that UHECRs can be accelerated to energies beyond $10^{20}$ eV through shocks, turbulence, and relativistic shear in jet-induced flows of Fanaroff-Riley (FR) type RGs. The resulting time-asymptotic UHECR spectrum is well modeled by a double power law with an ``extended'' exponential cutoff, primarily shaped by relativistic shear acceleration. In this study, we adopt this novel source spectrum and simulate the propagation of UHECRs from nearby RGs using the CRPropa code. We focus on Virgo A, Centaurus A, Fornax A, and Cygnus A, expected to be the most prominent UHECR sources among RGs. We then analyze the energy spectrum and mass composition of UHECRs arriving at Earth. We find that, due to the extended high-energy tail in the source spectrum, UHECRs from Virgo A, which has a higher Lorentz factor, exhibit a higher flux at the highest energies and a lighter mass composition at Earth, compared to those from Centaurus A and Fornax A with lower Lorentz factors. Despite Cygnus A having an even higher Lorentz factor, the large distance limits its contribution. With a small number of nearby prominent RGs, our findings suggest that if RGs are the major sources of UHECRs, the energy spectrum and mass composition of observed UHECRs would exhibit hemispheric differences between the Northern and Southern skies at the highest energies.

\end{abstract}


\section{Introduction}\label{s1}

Ultra-high-energy cosmic rays (UHECRs), with energies exceeding $1~{\rm EeV}(=10^{18}$ eV), remain one of the enigmatic phenomena in astrophysics. Despite significant progress, their origins are not yet fully understood, making source identification an ongoing challenge. Theoretical models for their production invoke a variety of acceleration mechanisms, including diffusive shock acceleration (DSA) at both nonrelativistic and relativistic shocks, relativistic shear acceleration, turbulent acceleration, magnetic reconnection, and unipolar induction in the magnetospheres of neutron stars and black holes \citep[see][for recent comprehensive reviews]{anchordoqui2019, batista2019, coleman2023}. For instance, shock waves in extragalactic objects such as active galactic nuclei (AGNs), starburst galaxies, gamma-ray bursts (GRBs), and galaxy clusters have been studied in detail as potential accelerators \citep[e.g.,][]{honda2009, waxman1995, kang1996, anchordoqui1999}.

A key constraint on the accelerator of UHECRs is the Hillas criterion, which sets the maximum energy of a cosmic ray (CR) being confined within an astrophysical source of size $\mathcal{R}$ and magnetic field strength $\mathcal{B}$:
\begin{equation}
E_{H} \approx 0.9~{\rm EeV} \cdot Z \left(\frac{\mathcal{B}}{1~\muG}\right) \left(\frac{\mathcal{R}}{1~\rm{kpc}}\right),
\label{Ehillas}
\end{equation}
where $Z$ is the charge number of the CR particle. This geometrical constraint provides a first-order estimate of the maximum energy achievable before particles escape the system \citep{hillas1984}. Relativistic jets of Fanaroff–Riley (FR) type radio galaxies (RGs), with $\mathcal{R}\sim$ a few $-100$ kpc and $\mathcal{B}\sim 10-100 \muG$, can yield $E_{H}\gtrsim10^{20}$ eV, making them as promising sites of UHECR acceleration \citep[e.g.,][]{eichmann2018, fang2018, kimura2018, rieger2019, rachen2019, matthews2020, rodrigues2021, eichmann2022}.

FR-type RGs are generally classified into two categories based on the radio morphology of their jets \citep{fanaroff1974}: FR-I jets exhibit central brightening, while FR-II jets are edge-brightened. Typically, FR-I jets are associated with lower radio luminosities, indicative of reduced kinetic power, compared to FR-II jets \citep[e.g.,][]{godfrey2013}. Additionally, FR-I jets tend to decollimate and decelerate to subrelativistic speeds at kpc scales, whereas FR-II jets extend to distances of $\gtrsim 100$ kpc \citep[e.g.,][]{bicknell1995,laing2014,mingo2019}. These distinctions in jet morphology and dynamics are thought to be primarily governed by the jet power, $Q_j$ \citep[e.g.,][]{kaiser1997,perucho2007,godfrey2013}, and {\color{blue}the jet-to-background density contrast, $\eta=\rho_j/\rho_b$} \citep[e.g.,][]{perucho2014,hardcastle2020}. A study by \citet{bhattacharjee2024} further demonstrated that in fact the initial bulk Lorentz factor of the jet, $\Gamma_j \equiv 1/\sqrt{1-\beta_j^2} \propto (Q_j/\eta\pi r_j^2)^{1/2}$ where $r_j$ is the radius of the jet inflow, and the jet-head advance speed, $v_{\rm head}^*/c \propto \eta^{1/2}\Gamma_j \propto (Q_j/\pi r_j^2)^{1/2}$, are crucial parameters governing the dynamics of jet flows. Here, $\beta_j\equiv v_j/c$ is the initial jet speed.

The acceleration processes of CRs that operate in relativistic jets have been extensively investigated. These include DSA at subrelativistic shocks \citep[e.g.,][]{matthews2019} and turbulent shear acceleration (TSA) \citep[e.g.,][]{hardcastle2010,ohira2013} that operate within the backflow (jet's cocoon). Gradual shear acceleration (GSA) occurs mainly in jet-spine and backflow regions \citep[e.g.,][]{webb2018,rieger2019}, while non-gradual shear acceleration (nGSA) takes place at the interface between the jet-spine and backflow \citep[e.g.,][]{ostrowski1998,caprioli2015,kimura2018}. The term relativistic shear acceleration (RSA) is commonly used to encompass both GSA and nGSA.

Recently, using relativistic hydrodynamic (RHD) simulations to model jet dynamics, in combination with Monte Carlo simulations for the transport of CRs, we investigated the acceleration of CRs in FR-I and FR-II jets \citep{seo2023,seo2024}. Owing to limitations in numerical resolution and the hydrodynamic framework, magnetic field distributions and magnetohydrodynamic (MHD) turbulence were incorporated using physically motivated prescriptions. Then, along with the flow dynamics of jets from simulations, we tracked the scattering of CRs in fluctuating magnetic fields and their subsequent energization. The study revealed that for CRs with energies $E\lesssim1$ EeV, DSA is the dominant energy-gain process. In contrast, for higher energies ($E\gtrsim1$ EeV), RSA becomes the primary acceleration mechanism, enabling particles to reach energies exceeding $10^{20}$ eV. TSA is active across the entire energy range, but its contribution is comparatively less significant (see Figure 5 of \citet{seo2024}).

We also found that the time-asymptotic energy spectrum of UHECRs escaping from the jet system can be approximated by a double power law (DPL) with an exponential cutoff (see Equation (\ref{DPLexp}) below). Below the break energy, $E_b$, the spectral index ranges from $\sim-0.5$ to $-0.6$, while above $E_b$, it steepens to $\sim-2.6$. The cutoff is ``extended'' as $E_b \langle \Gamma \rangle_{\rm spine}^2$, modulated by the square of $\langle \Gamma \rangle_{\rm spine}$, the mean Lorentz factor of the jet spine. On the other hand, previous studies have conventionally assumed a single power law (SPL) with a cutoff for the injection spectrum at RGs (see Equation (\ref{SPL}) below) \citep[e.g.][]{PAO2017JCAP,eichmann2018,eichmann2022,batista2019a}. With the SPL model, a power-law index of $\sim -2.0$ to $-2.3$ has been commonly adopted, based on DSA at nonrelativistic and relativistic shocks. However, analyses of observed UHECR spectra (see below) have suggested a broad range for the source power-law index, $\sim-1$ to $-2$.

Within the so-called GZK horizon ($\sim 100$ Mpc), which represents the attenuation limit of UHECRs due to interactions with cosmic microwave background (CMB) photons \citep{greisen1966, zatsepin1966}, about forty FR RGs have been identified \citep{velzen2012,rachen2019}. Among the nearest RGs, Virgo A (M87) and Centaurus A exhibit blazar-like properties. The inner jet of Virgo A appears one-sided, suggesting bulk relativistic motion; it shows superluminal motions observed in optical and X-ray, indicating an mean Lorentz factor along the jet-spine, $\langle\Gamma\rangle_{\rm{spine}}\gtrsim6$ \citep[e.g.,][]{biretta1999,snios2019}. Similarly, Centaurus A features a one-sided jet, albeit with lower velocities, which is interpreted to have $\langle\Gamma\rangle_{\rm{spine}}\sim1.2-2$ \citep[e.g.,][]{wykes2019,snios2019b}. In contrast, Fornax A hosts two inner jets associated with recent low-power activity, along with expansive lobes in its outer regions \citep[e.g.,][]{geldzahler1984,maccagni2020}. These nearby prominent RGs are considered potential major contributors to observed UHECRs \citep{romero1996,eichmann2018,matthews2018,kim2019,eichmann2022,wang2024}.

Currently, two leading experiments dedicated to observing UHECRs are in operation: the Pierre Auger Observatory (PAO) in Argentina, which primarily covers the Southern Hemisphere \citep[e.g.,][]{PAO2015}, and the Telescope Array (TA) in the United States, which focuses on the Northern Hemisphere \citep[e.g.,][]{TA2008}. Despite their shared objectives and similar experimental techniques, the two experiments report notable discrepancies in their observations. In particular, the particle flux measured by PAO is consistently lower than that observed by TA \citep{aab2020, tsunesada2022, kim2023}, and the inferred mass composition, based on air-shower maximum ($X_{\rm max}$) measurements, appears heavier at PAO than at TA \citep{PAO2024, zhezher2022}.

To reconcile the discrepancies, PAO-TA working groups have conducted joint analyses of their data. The energy spectrum working group demonstrated that by adjusting the CR energy scale of PAO by $+4.5\%$ and that of TA by $-4.5\%$, the spectra from both experiments align well at energies $E\lesssim10^{19.5}$ eV \citep{tsunesada2022, tsunesada2023}. However, a substantial difference persists at $E\gtrsim10^{19.5}$ eV, though the difference is smaller in the spectra within the overlapping sky region of the two experiments. Furthermore, after accounting for systematic effects in both experiments and employing the same hadronic interaction, the mass composition working group found no significant differences in the mean depth of air-shower maximum, $\langle X_{\rm max} \rangle$, between the Auger and TA measurements at $E\lesssim10^{19.5}$ eV \citep{PAO2024icrc}. However, the agreement in the standard deviation, $\sigma(X_{\rm max})$, remains unclear. In addition, the comparison analysis of mass composition could not be made for higher energies of $E\gtrsim10^{19.5}$ eV due to the limitation of data availability. Thus, the asymmetry in the observed UHECR properties between the Northern and Southern Hemispheres is not yet fully resolved, necessitating further investigation to confirm or refute its significance.

The search for the origins of UHECRs has long focused on detecting anisotropy in their arrival directions. For example, PAO reported evidence of a large-scale dipole anisotropy at $E\gtrsim8$ EeV, which roughly correlates with the local distribution of galaxies \citep{PAO2007,PAO2017}. TA, on the other hand, identified localized ``hot spots'' of excess UHECR events \citep{abbasi2014}. While these could offer crucial insights into their astrophysical sources, the arrival directions of UHECRs are altered due to deflection caused by extragalactic and Galactic magnetic fields as they propagate from extragalactic sources to Earth \citep[e.g.,][]{sigle2004, takami2008, das2008, Ryu2010, hackstein2018, Korochkin2025, Rossoni2025}. Hence, the eventual identification of the sources requires a detailed understanding of the intervening magnetic fields. Moreover, the deflection increases travel distances and induces substantial time delays relative to rectilinear trajectories, further complicating source identification.

FR-type RGs are plausible sources of UHECRs, as described above. \citet{eichmann2022}, for instance, employing a SPL model with an exponential cutoff for the source spectrum, investigated the energy spectrum, mass composition, and dipole and quadrupole anisotropies produced by nearby prominent RGs, along with isotropic contributions from low-luminosity RGs. However, our newly proposed DPL source spectrum with an ``extended'' exponential cutoff could significantly impact the energy spectrum and mass composition of UHECRs arriving at Earth, while having minimal effect on their arrival direction distribution. Building on these insights, we investigate how our proposed DPL source spectrum (Equation (\ref{DPLexp}) below) results in the energy spectrum and mass composition of observed UHECRs, compared to the conventional SPL source spectrum. To this end, we perform Monte Carlo simulations of UHECR propagation from various relativistic jet models linked with nearby prominent RGs, using the publicly available CRPropa code \citep{batista2016, batista2022}. Our study examines the influence of key model parameters, such as $E_b$ and $\langle \Gamma \rangle_{\rm spine}$, in the source spectrum, the metallicity of the interstellar medium (ISM) in host galaxies, and extragalactic background light (EBL) models, and the effects of extended propagation distances due to deflections by intervening magnetic fields.

This paper is organized as follows: Section \ref{s2} describes the simulation setup and model parameters. Section \ref{s3} presents the results, including the energy spectrum and mass composition of UHECRs reaching Earth. A summary is provided in Section \ref{s4}.

\begin{figure}[t]
\vskip 0.2 cm
\hskip -0.6 cm
\includegraphics[width=1.1\linewidth]{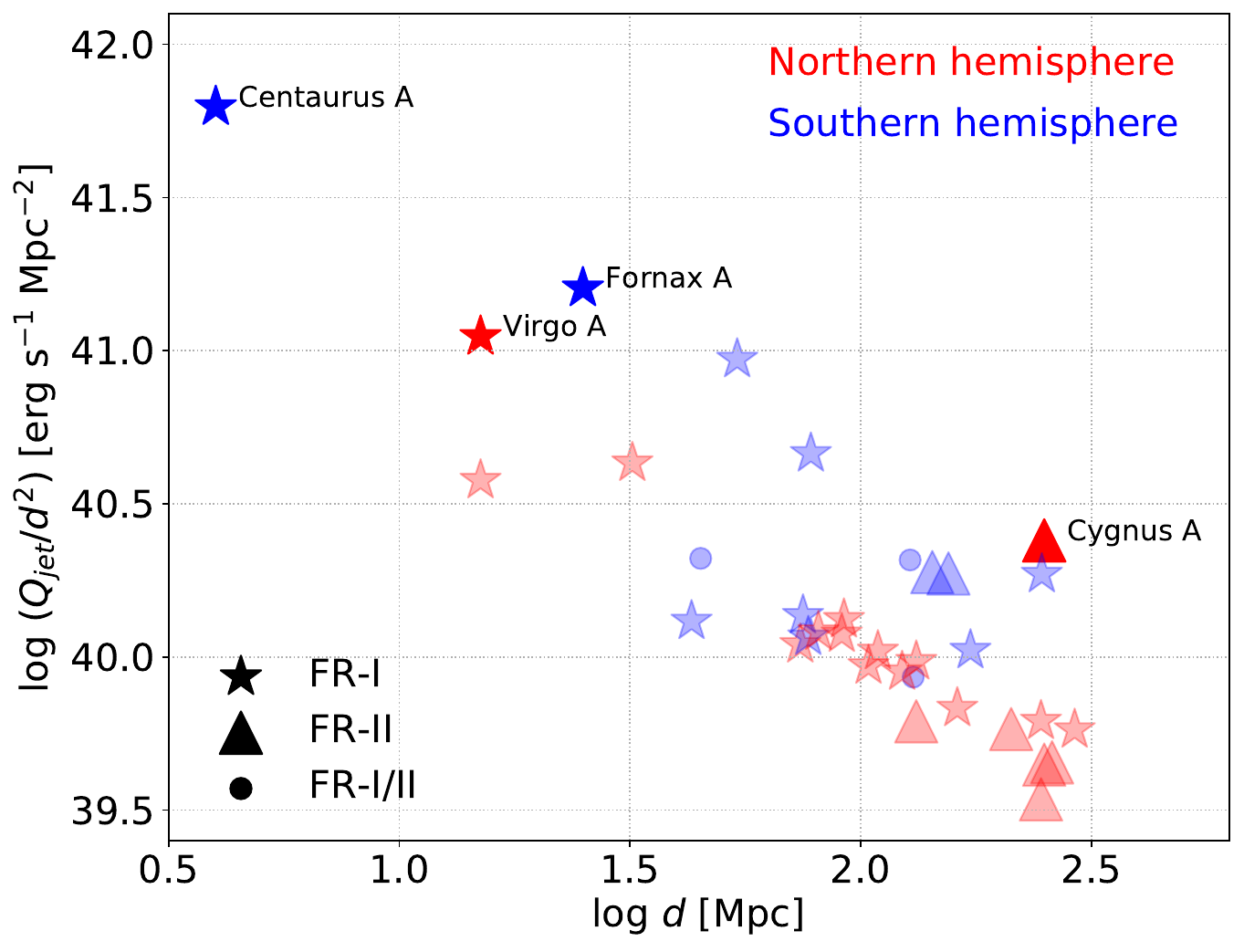}
\vskip -0.2 cm
\caption{Jet kinetic power flux, $Q_j /d^2$, for 42 local RGs as a function of distance $d$. Different symbols distinguish between FR-I, FR-II, and FR-I/II types, with FR-I/II exhibiting characteristics intermediate between FR-I and FR-II types. Red and blue symbols represent RGs located in the Northern and Southern Hemispheres, respectively. Thick symbols highlight the four prominent RGs listed in Table \ref{t1}.}
\label{fig1}
\end{figure}

\section{Models and Simulations}\label{s2}

Figure \ref{fig1} illustrates the jet kinetic power flux, $Q_j/d^2$, as a function of distance $d$ for 42 local RGs, compiled from a catalog in the literature \citep{velzen2012,rachen2019}. This represents a virtually complete sample of RGs within 300 Mpc. The figure suggests that Centaurus A, Virgo A, and Fornax A are likely the primary contributors to UHECRs detected on Earth \citep[see also][]{eichmann2018,eichmann2022}. Here, we focus on these three RGs along with Cygnus A. Despite lower $Q_j/d^2$ due to its large distance, Cygnus A, an FR-II RG, is included because it is one of the brightest radio sources in the sky and has a very high jet Lorentz factor. Table \ref{t1} lists the jet parameters for four prominent RGs, including the jet power $Q_j$, the mean Lorentz factor of the jet spine $\langle \Gamma \rangle_{\rm spine}$, and distance $d$, as inferred from various observations cited in the references.

\begin{deluxetable*}{ccccccccc}[ht]
\tablecaption{Jet Parameters of Prominent Radio Galaxies Inferred from Observations\label{t1}}
\tabletypesize{\scriptsize}
\tablecolumns{7}
\tablenum{1}
\tablewidth{0pt}
\tablehead{
\colhead{Name} &  
\colhead{$Q_j$}&
\colhead{$\varphi \cdot \xi$} &
\colhead{$E_b$} &
\colhead{$\langle\Gamma\rangle_{\rm{spine}}$} &
\colhead{FR type} &
\colhead{$d$} &
\colhead{references}\\
\colhead{} & 
\colhead{($\rm erg~s^{-1}$)} &
\colhead{} &
\colhead{(EeV)} &
\colhead{} &
\colhead{} &
\colhead{(Mpc)} &
\colhead{} }

\startdata
Virgo A     & $\sim 2.5\times 10^{43}$ & $\sim0.45$ & $\sim10$   & $\sim7.0$  & FR-I  & $\sim16$  & \citet{biretta1999}, \citet{kino2022} \\
Centaurus A & $\sim1.0\times 10^{43}$ & $\sim0.15$ & $\sim3.0$  & $\sim1.2$  & FR-I  & $\sim4.0$ & \citet{snios2019}, \citet{wykes2019} \\
Fornax A    & $\sim1.0\times 10^{44}$ & $\sim0.2$  & $\sim7.0$  & $\sim1.5$  & FR-I  & $\sim25$  & \citet{geldzahler1984}, \citet{lanz2010}\\
Cygnus A    & $\sim1.5\times 10^{45}$ & $\sim0.7$  & $\sim50$   & $\sim10$   & FR-II & $\sim250$ & \citet{godfrey2013}\\
\hline
\hline
\enddata
\end{deluxetable*}

\begin{deluxetable*}{ccccccc}[ht]
\tablecaption{Jet Models with Parameter Variations \label{t2}}
\tabletypesize{\small}
\tablecolumns{7}
\tablenum{2}
\tablewidth{0pt}
\tablehead{
\colhead{Model name$^a$} &
\colhead{$E_{\rm max}$ (EeV)}&
\colhead{$\gamma$} &
\colhead{$d$ (Mpc)} &
\colhead{Metallicity$^c$} &
\colhead{EBL model$^d$} &
\colhead{$d_p$ range$^e$}\\
\colhead{(1)}& \colhead{(2)} & \colhead{(3)} & \colhead{(4)} & \colhead{(5)} & \colhead{(6)} & \colhead{(7)}
}
\startdata
SPL1 &$50$   & $-1$  & 16 & $3\times f_{\sun}$ & Gilmore et al.& $1-1.5 d$\\
SPL2 &$100$  & $-2$  & 16 & $3\times f_{\sun}$ & Gilmore et al.& $1-1.5 d$\\
\hline
\hline
{Model name$^b$} & {$E_b$ (EeV)}& {$\langle\Gamma\rangle_{\rm{spine}}$} & {$d$ (Mpc)} & {Metallicity$^c$} & {EBL model$^d$} &{$d_p$ range$^e$}\\
\hline
VirA1& 10  & 7.0  & 16 & $3\times f_{\sun}$ & Gilmore et al. & $1-1.5 d$\\
VirA2 & 30  & 7.0  & 16 & $3\times f_{\sun}$ & Gilmore et al. & $1-1.5 d$\\
VirA3 & 10  & 3.0  & 16 & $3\times f_{\sun}$ & Gilmore et al. & $1-1.5 d$\\
VirA4 & 10  & 7.0  & 16 & $1\times f_{\sun}$ & Gilmore et al. & $1-1.5 d$\\
VirA5 & 10  & 7.0  & 16 & $3\times f_{\sun}$ & Dom\'inguez et al. & $1-1.5 d$\\
VirA6 & 10  & 7.0  & 16 & $3\times f_{\sun}$ & Gilmore et al. & $1-3 d$\\
\hline
CenA1 & 3.0  & 1.2  & 4.0& $3\times f_{\sun}$ & Gilmore et al.& $1-1.5 d$\\
CenA2 & 6.0  & 1.2  & 4.0& $3\times f_{\sun}$ & Gilmore et al.& $1-1.5 d$\\
CenA3 & 3.0  & 1.2  & 4.0& $1\times f_{\sun}$ & Gilmore et al.& $1-1.5 d$\\
\hline
ForA1 & 7.0  & 1.5  & 25& $3\times f_{\sun}$ & Gilmore et al.& $1-1.5 d$\\
ForA2 & 15   & 1.5  & 25& $3\times f_{\sun}$ & Gilmore et al.& $1-1.5 d$\\
ForA3 & 7.0  & 1.5  & 25& $3\times f_{\sun}$ & Dom\'inguez et al.& $1-1.5 d$\\
\hline
CygA1& 50  & 10  & 250& $3\times f_{\sun}$ & Gilmore et al. & $1-1.5 d$\\
CygA2& 50  & 10  & 250& $3\times f_{\sun}$ & Dom\'inguez et al. & $1-1.5 d$\\
\hline
\hline
\enddata
\tablenotetext{a}{Single power-law models given in Equation (\ref{SPL}) with $E_{\rm max}$ and $\gamma$ as parameters.}
\tablenotetext{b}{Double power-law models given in Equation (\ref{DPLexp}) with $E_b$ and $\langle\Gamma\rangle_{\rm{spine}}$ as parameters.}
\tablenotetext{c}{Abundance of nuclei heavier than helium in the ISM of host galaxies.}
\tablenotetext{d}{Extragalactic background light models from \citet{gilmore2012} or \citet{dominguez2011}.}
\tablenotetext{e}{Propagation distance, $d_p$, reflecting possible deflection of paths due to intervening magnetic fields.}
\vskip -0.7 cm
\end{deluxetable*}

\subsection{Energy Spectrum of UHECRs from RGs}\label{s2.1}

As mentioned in the introduction, we showed that the energy spectrum of UHECRs from relativistic jets can be approximated by a DPL with an ``extended'' exponential cutoff \citep{seo2023, seo2024}. Based on it, in this work, we model the energy spectrum of UHECRs emitted per unit time from a jet with $Q_j$ and $\langle\Gamma\rangle_{\rm{spine}}$ as
\begin{equation}
\begin{aligned}
\frac{d{N}}{dE_0 dt}=S_n f_r(A_0)\left(\left(\frac{E_0}{Z_0 E_b}\right)^{-s_1}+\left(\frac{E_0}{Z_0 E_b}\right)^{-s_2}\right)^{-1}\\
\times\exp\left(-\frac{E_0}{Z_0 E_b \langle\Gamma\rangle_{\rm{spine}}^2}\right),~~~~~~~~~~~~~~~~
\label{DPLexp}
\end{aligned}
\end{equation}
where $E_0$, $Z_0$, and $A_0$ are the energy, charge number, and atomic mass of UHECRs, respectively, with the subscript ``0'' referring to source quantities. Here, $Z_0 E_b$ is the ``rigidity-dependent'' break energy, and $E_b(Q_j)$ is primarily determined by $Q_j$ but also depends on other jet properties (see below). The power-law indices are approximately $s_1\approx -0.6$ for FR-I jets and $s_1\approx -0.5$ for FR-II jets, while $s_2\approx-2.6$ for both types of jets. The extension factor $\langle\Gamma\rangle_{\rm{spine}}^2$ in the exponential cutoff is included, accounting for the energy boost via relativistic non-gradual shear acceleration. The normalization factor $S_n$ may be determined by fixing the total power of UHECRs, or the luminosity of UHECRs, emitted from the jet, $L_{\rm CR}= \int (dN/dE_0 dt)E_0 dE_0$; for simplicity, we assume $L_{\rm CR} \propto Q_j$. Finally, $f_r (A_0)$ represents the fraction of CR particles with $A_0$.

The determination of the key parameters, $E_b$ and $\langle\Gamma\rangle_{\rm{spine}}$, from observations of RGs is far from straightforward and necessarily involves modeling. \citet{seo2023,seo2024} demonstrated through Monte Carlo simulations that the break energy $E_b$ arises from the interplay among particle acceleration via DSA and RSA, confinement due to particle scattering off background MHD turbulence, and escape processes within the jet's cocoon. While higher-energy particles are primarily accelerated via RSA near the jet-spine flow, the value of $E_b$ is governed by confinement and escape processes that occur over the broader cocoon region, as shown in Figure 6 of \citet{seo2024}. In the regime of efficient acceleration, $E_b$ approaches the Hillas energy, $E_b\approx E_H \propto \mathcal{B}\cdot W_c$, where the system size is set by the cocoon width, $W_c$. By contrast, under inefficient (slow) acceleration, particles escape more readily through rapid diffusion, leading to a reduced break energy, $E_b \approx \varphi E_H$, where $\varphi\approx {\rm min}[1, \langle \Gamma\beta\rangle_{\rm acc}] \sim 0.4\text{–}1$. This dimensionless factor encapsulates the competition between acceleration and confinement within the cocoon and is inferred from the simulations.  The value of $\langle \Gamma\beta\rangle_{\rm acc}$ can be estimated along the simulated jet-spine velocity profile.

Our Monte Carlo simulations also showed that the Hillas energy $E_H$ is mainly governed by the jet power and scales as $E_H\propto \xi Q_j^{\alpha}$ with $\alpha=1/4$ for FR-I jets and $\alpha=1/3$ for FR-II jets. Here, $\xi \propto r_j^{1/2}\rho_b^{1/4}\sim 0.4-1$ is another numerically derived factor with only a weak dependence on jet size and ambient medium density. Accordingly, we adopt a model in which the break energy is primarily determined by the jet power:
\begin{equation}
E_b\approx 45~{\rm EeV} \times \varphi \cdot \xi \cdot \left(\frac{Q_j}{Q_n}\right)^{\alpha}, 
\label{ebreak}
\end{equation}
where $Q_n=3.5\times 10^{44}~{\rm erg s^{-1}}$. The combined factor, $\varphi \cdot \xi$, tends to be larger for FR-II jets compared to FR-I jets. For the radio galaxies listed in Table \ref{t1}, we adopt a representative range of $\varphi \cdot \xi \sim 0.15-0.7$, while acknowledging some degree of arbitrariness in these values. Given the uncertainties in our modeling of turbulent dynamo and particle scattering, any further refinement of the $\varphi \cdot \xi$ range would be speculative and unwarranted at this stage of theoretical development. For further details of the $E_b$ modeling, see \citet{seo2023, seo2024}. In the next section, we examine how variations in $\varphi \cdot \xi$ influence our results.

The initial Lorentz factor, $\Gamma_j \propto (Q_j / \eta \pi r_j^2)^{1/2}$, is proportional to the jet kinetic power flux per unit mass \citep{bhattacharjee2024}. Here, again, $\eta=\rho_j/\rho_b$ denotes the jet-to-background density contrast. RHD simulations of FR jets, as presented in \citet{seo2024} and \citet{bhattacharjee2024}, indicate that the mean Lorentz factor of the jet spine, $\langle \Gamma \rangle_{\rm spine}$, at the fully developed stage of the jet also depends on the extent of decollimation and deceleration of the jet-spine flow and may not follow the same scaling as $\Gamma_j$. Less powerful jets with lower $Q_j$ (and thus smaller $\Gamma_j$) experience more significant deceleration, resulting in smaller $\langle \Gamma \rangle_{\rm spine}$ and a less pronounced extended exponential tail in the energy spectrum.
   
Superluminal motion observed in some relativistic jets provides a practical means to estimate $\langle \Gamma \rangle_{\rm spine}$ \citep[e.g.,][]{Lister2009}. Therefore, for zeroth-order modeling, we adopt representative values of $\langle \Gamma \rangle_{\rm spine}$ in Table \ref{t1}, derived from observational data on kpc scales, as cited in the corresponding references. For jets such as Virgo A and Fornax A, which exhibit a compact inner jet and extended outer lobes, particle acceleration processes are expected to be primarily governed by jet conditions on kpc scales. In the following section, we explore how variations in $\langle \Gamma \rangle_{\rm spine}$ affect our results.

As noted in the introduction, previous studies have typically modeled the energy spectrum of UHECRs from RGs as a SPL with a cutoff, expressed as $dN/dE_0 \propto E_0^{\gamma}F_{\rm cut}(E_0/Z_0 E_{\rm max})$. The cutoff function $F_{\rm cut}$ sharply decreases above the rigidity-dependent maximum energy, $Z_0 E_{\rm max}$, where $E_{\rm max}$ is the maximum energy that protons can attain in the source. Hence, to provide a comparison to the spectrum in Equation (\ref{DPLexp}), we also consider a SPL spectrum with an exponential cutoff given as\footnote{The minus sign is omitted from the power-law index to remain consistent with the convention used in Equation (\ref{DPLexp}), although its inclusion is more common in the literature.}:
\begin{equation}
\frac{d{N}}{dE_0 dt}=S_n f_r(A_0) \left(\frac{E_0}{Z_0 E_{\rm max}}\right)^{\gamma}\exp\left(-\frac{E_0}{Z_0 E_{\rm max}}\right).
\label{SPL}
\end{equation}
The power-law index in the range of $\gamma \sim -1$ to $-2$ is adopted, as this best reproduces observed UHECR spectra \citep[e.g.][]{PAO2017JCAP,batista2019a}.

\subsection{Jet Models}\label{s2.2}

To assess the energy spectrum and mass composition of UHECRs arriving at Earth from the four prominent RGs listed in Table \ref{t1}, we examine a number of jet models with DPL listed in Table \ref{t2}; the key jet parameters, $E_b$ and $\langle \Gamma \rangle_{\rm spine}$ in Equation (\ref{DPLexp}), are varied to evaluate their influence. Column (1) of Table \ref{t2} shows the model names: VirA, CenA, ForA, and CygA correspond to Virgo A, Centaurus A, Fornax A, and Cygnus A, respectively. The models labeled with ``1'' serve as the representative cases with the parameter values given in Table \ref{t1}, while those with other numbers denote variants. For instance, VirA1 is the representative model for Virgo A, whereas VirA2 has a twice higher $E_b$ and VirA3 has a lower $\langle \Gamma \rangle_{\rm spine}$. Additionally, we include two SPL models with different values of $E_{\rm max}$ and $\gamma$ in Equation (\ref{SPL}). They are intended to serve as comparison models for VirA1, and hence, the normalization $S_n$ is fixed such that $dN/dE_0$ matches that of VirA1 at 1 EeV.

\begin{figure}[t]
\vskip 0.2 cm
\hskip -0.2 cm
\includegraphics[width=\linewidth]{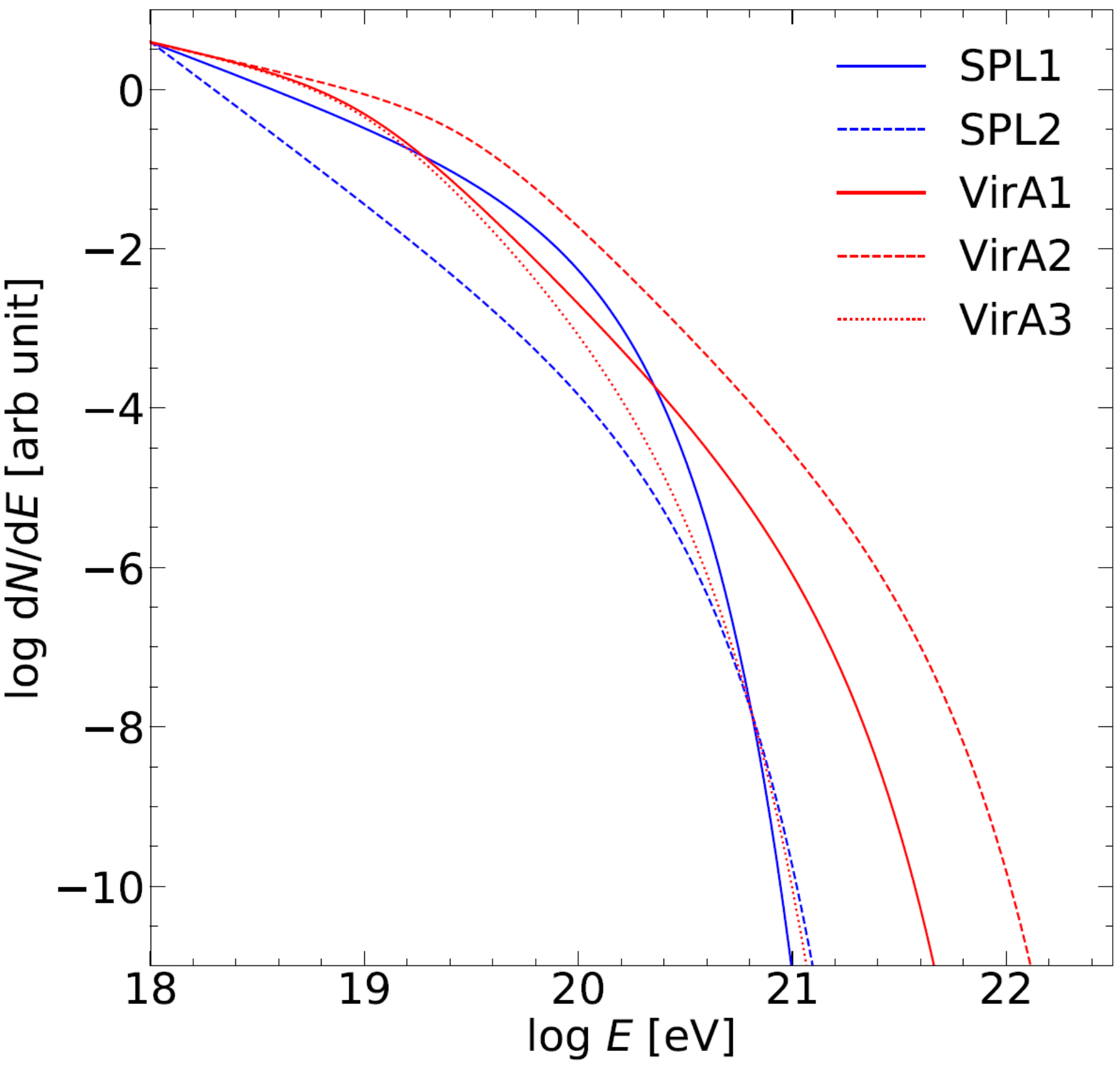}
\vskip -0.1 cm
\caption{Model energy spectra of UHECRs from RGs. The blue solid and dashed lines represent the single power-law models, SPL1 and SPL2, respectively. The red solid, dashed, and dotted lines correspond to the double power-law models, VirA1, VirA2, and VirA3, respectively, for Virgo A. See Table \ref{t2} for model parameters. Here, for comparison, the amplitude normalization is adjusted so that the spectra are aligned at $10^{18}$ eV across all models.}
\label{fig2}
\end{figure}

Figure \ref{fig2} presents the source spectra of SPL1, SPL2, VirA1, VirA2, and VirA3. Here, for illustration purposes only, $S_n$ is adjusted so that the spectra of all the five models align at 1 EeV. Although the SPL models have $E_{\rm max}$ greater than $E_b$ of the DPL models, VirA1 and VirA2, which have a high $\langle \Gamma \rangle_{\rm spine}$ value and thus a significantly extended exponential cutoff, exhibit pronounced high-energy tails. VirA2 shows an even more pronounced tail due to its higher $E_b$. On the other hand, VirA3, with a lower $\langle \Gamma \rangle_{\rm spine}$, displays a much less distinct high-energy tail, compared to VirA1 and VirA2.

\subsection{Metallicity in Host Galaxies}\label{s2.3}

Many RGs, including those in Table 1, are hosted by elliptical galaxies or galaxies with enhanced star-formation activity \citep{matthews1964,comerford2020}. These host galaxies generally exhibit higher metallicity than spiral galaxies like the Milky Way \citep{henry1999, pipino2008, mao2018}. Hence, for the fraction of UHECRs with atomic mass $A_0$, $f_r(A_0)$ in Equations (\ref{DPLexp}) and (\ref{SPL}), we adopt the following as fiducial values: 
\begin{equation}
\begin{array}{r@{}l}
f_r (A_0=1)=0.71, \\
f_r (2\le A_0 \le4)=0.21, \\
f_r (5 \le A_0 \le 22)=0.064, \\
f_r (23 \le A_0 \le 38)=0.0011, \\
f_r (39 \le A_0 \le 56))=0.0054,
\label{metallicity}
\end{array}
\end{equation}
presented for five mass groups of UHECRs. These values increase the abundance of nuclei heavier than helium by a factor of three compared to the Galactic CR abundance, as commonly assumed in previous studies \citep[e.g.,][]{kimura2018}. This fiducial metallicity is referred to as ``$3\times f_{\sun}$'' in column (5) of Table \ref{t2}. To explore the effects of different metallicities, we also consider VirA4 and CenA3 using ``$1\times f_{\sun}$'', where the abundance of nuclei heavier than helium matches the Galactic CR abundance.

\subsection{Propagation of CRs}\label{s2.4}

As UHECRs travel through the intergalactic and Galactic space to Earth, they interact with CMB photons, leading to the GZK effect, and also with EBL photons \citep[e.g.][]{allard2012}, while their trajectories are deflected by intervening magnetic fields. The propagation of UHECRs from our jet models in Table \ref{t2} is simulated using the CRPropa code, which incorporates a comprehensive suite of interactions and energy-loss processes relevant for UHECR propagation \citep{batista2016, batista2022}. These include Bethe-Heitler pair production, photopion production, and photo-disintegration due to interactions with the CMB and EBL, as well as adiabatic energy loss and nuclear decay \citep{batista2022}. Since UHECR propagation could be sensitive to the choice of EBL model, we consider two widely used versions: the models by \citet{gilmore2012} and \citet{dominguez2011}. Specifically, while we adopt \citet{gilmore2012} as the fiducial EBL model, for three jet models, VirA5, ForA3, and CygA2, the \citet{dominguez2011} model is used, as indicated in column (6) of Table \ref{t2}.

\begin{figure*}[t]
\vskip 0.1 cm
\hskip 0.3 cm
\centering
\includegraphics[width=0.9\linewidth]{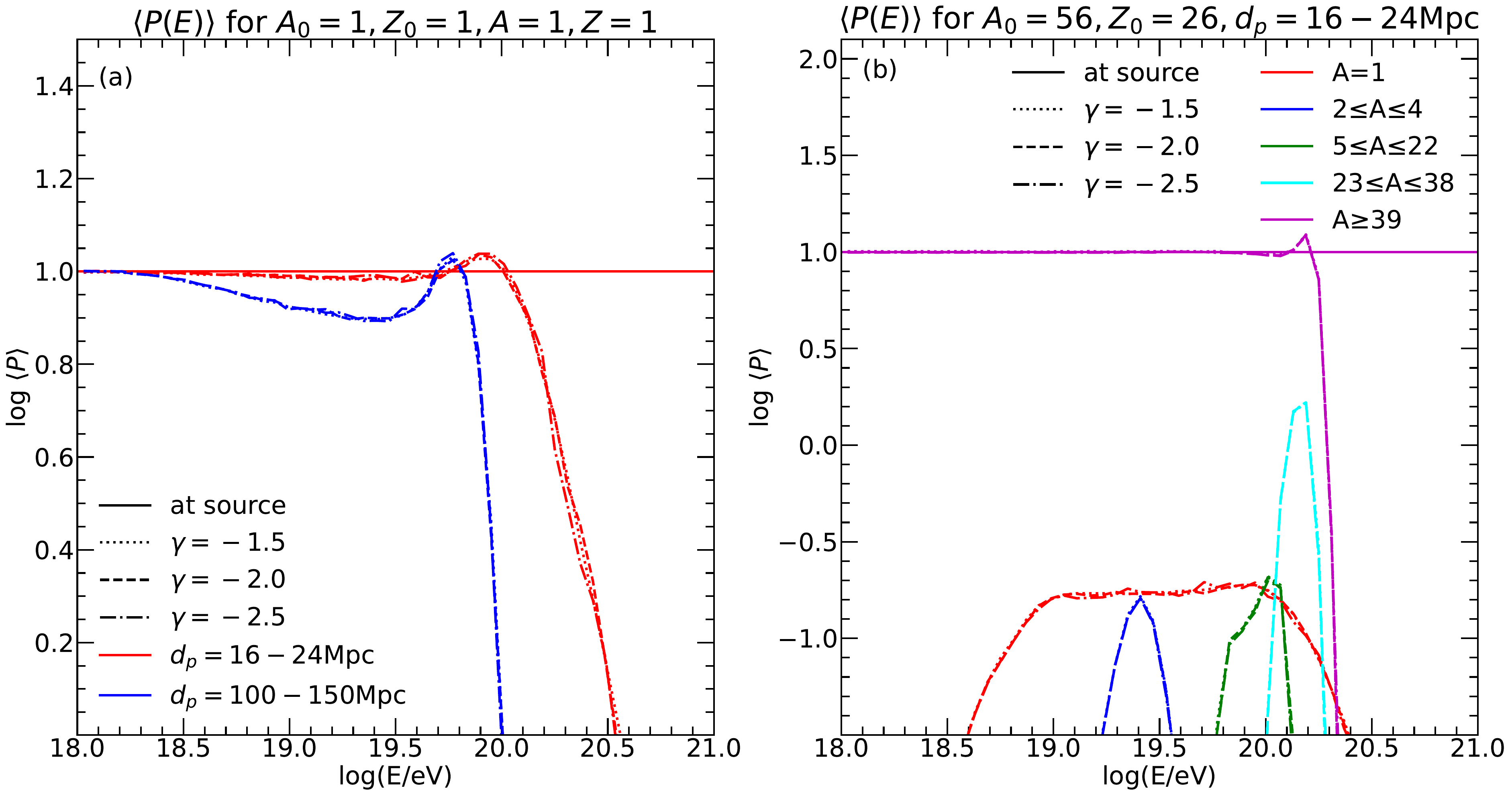}
\vskip -0.1 cm
\caption{Propagation function, $\langle P \rangle$, for a power-law injection spectrum with different spectral indices, $\gamma=-1.5$ (dotted lines), $-2.0$ (dashed lines), and $-2.5$ (dot-dashed lines). (a) $\langle P \rangle$ for protons ($A_0=1$, $Z_0=1$, $A=1$, and $Z=1$) over $d_p= 16-24$ Mpc (red) and $d_p = 100-150$ Mpc (blue). (b) $\langle P \rangle$ with ion nuclei at the source ($A_0=56$ and $Z_0=26$) over $d_p = 16-24$ Mpc. Different colors represent the propagation functions for five mass groups. For the EBL, the model of \citet{gilmore2012} is used.}
\label{fig3}
\vskip 0.1 cm
\end{figure*}

The deflection of UHECRs by intervening magnetic fields not only alters their arrival directions but also increases the path lengths depending on their rigidity and the strength and configuration of the magnetic fields \citep[e.g.,][]{das2008,takami2008,hackstein2016,garcia2021}. While the sky distribution of UHECRs originating from RGs and its anisotropy are not the scope of this paper, as noted in the introduction, the increase in travel distance and hence travel time can affect the energy spectrum and mass composition of UHECRs arriving at Earth. Given the limited knowledge of intervening intergalactic and Galactic magnetic fields, we take a practical approach by evaluating a statistical ensemble of one-dimensional (1D) propagation over distances spanning a range of values, rather than explicitly tracing three-dimensional (3D) trajectories influenced by the intervening magnetic fields. We adopt $d_p = (1 - 1.5)d$ as the fiducial model for the propagation distance range, but also consider $d_p = (1 - 3)d$ in the VirA6 model, as listed in column (7) of Table \ref{t2}. Here, $d$ represents the distance to the sources in Table \ref{t1}. For the CRPropa calculations, we assume a flat universe with cosmological parameters: $H_0 = 67.3~\mathrm{km~s^{-1}~Mpc^{-1}}$, $\Omega_m = 0.315$, and $\Omega_\Lambda = 0.685$.

The flux of UHECRs reaching Earth from a source with an injection energy spectrum $dN/dE_0dt$ at a distance $d$ may be expressed as
\begin{equation} 
\begin{aligned} 
F(E,A,Z)= \frac{1}{4\pi d^2} \frac{dN(E_0,A_0,Z_0)}{dE_0dt} \\
\times \langle P(E_0,A_0,Z_0,E,A,Z,d_p)\rangle, 
\end{aligned} 
\end{equation}
where $E_0$, $A_0$, and $Z_0$ refer to the quantities of UHECRs at the source, as mentioned above, while $E$, $A$, and $Z$ denote those of particles arriving at Earth. The propagation function, $\langle P(E_0,A_0,Z_0,E,A,Z,d_p)\rangle$, accounts for modification due to energy loss and disintegration of heavy nuclei during the journey through the intergalactic and Galactic space. The angle brackets are included to indicate that this function is averaged over randomly sampled propagation distances within the range $d_p$.

We calculate $\langle P(E_0,A_0,Z_0,E,A,Z,d_p)\rangle$ for a power-law injection spectrum, $dN/dE_0dt \propto f_r(A_0)(E_0/Z_0)^{\gamma}$ with $\gamma=-2$, with the metallicities in column (5) of Table \ref{t2}, using 1D CRPropa simulations with the EBL models in column (6), over the distance ranges in column (7). Although, in principle, the propagation function depends on the shape of the injection spectrum, we find that this dependence is relatively weak. To illustrate this, Figure \ref{fig3}(a) shows the propagation function for proton-only UHECR spectra ($A_0 = 1$, $Z_0 = 1$, $A = 1$, $Z = 1$) with three different power-law indices $\gamma$, over two different propagation distance ranges, $d_p = 16\times(1-1.5)$ Mpc (red) and $100\times(1-1.5)$ Mpc (blue). The figure confirms that the dependence on $\gamma$ is not strong. On the other hand, the energy loss is strongly influenced by the propagation distance. Figure \ref{fig3}(b) further illustrates this by presenting the propagation function for iron-only UHECR spectra at the source ($A_0 = 56$, $Z_0 = 26$) with three different $\gamma$ over $d_p = 16\times(1-1.5)$ Mpc. The figure includes the propagation functions for five mass groups of UHECRs arriving at Earth, illustrating the fragmentation of iron nuclei above $10^{20}$ eV into lighter species due to photo-disintegration. Again, the weak dependence on $\gamma$ is confirmed. In this work, we employ the propagation function for a power-law injection spectrum with $\gamma=-2$ and apply it to the SPL and DPL models in Table \ref{t2} to determine the energy spectrum of UHECRs reaching Earth.

\begin{figure*}[t]
\vskip 0 cm
\hskip 0 cm
\centering
\includegraphics[width=1\linewidth]{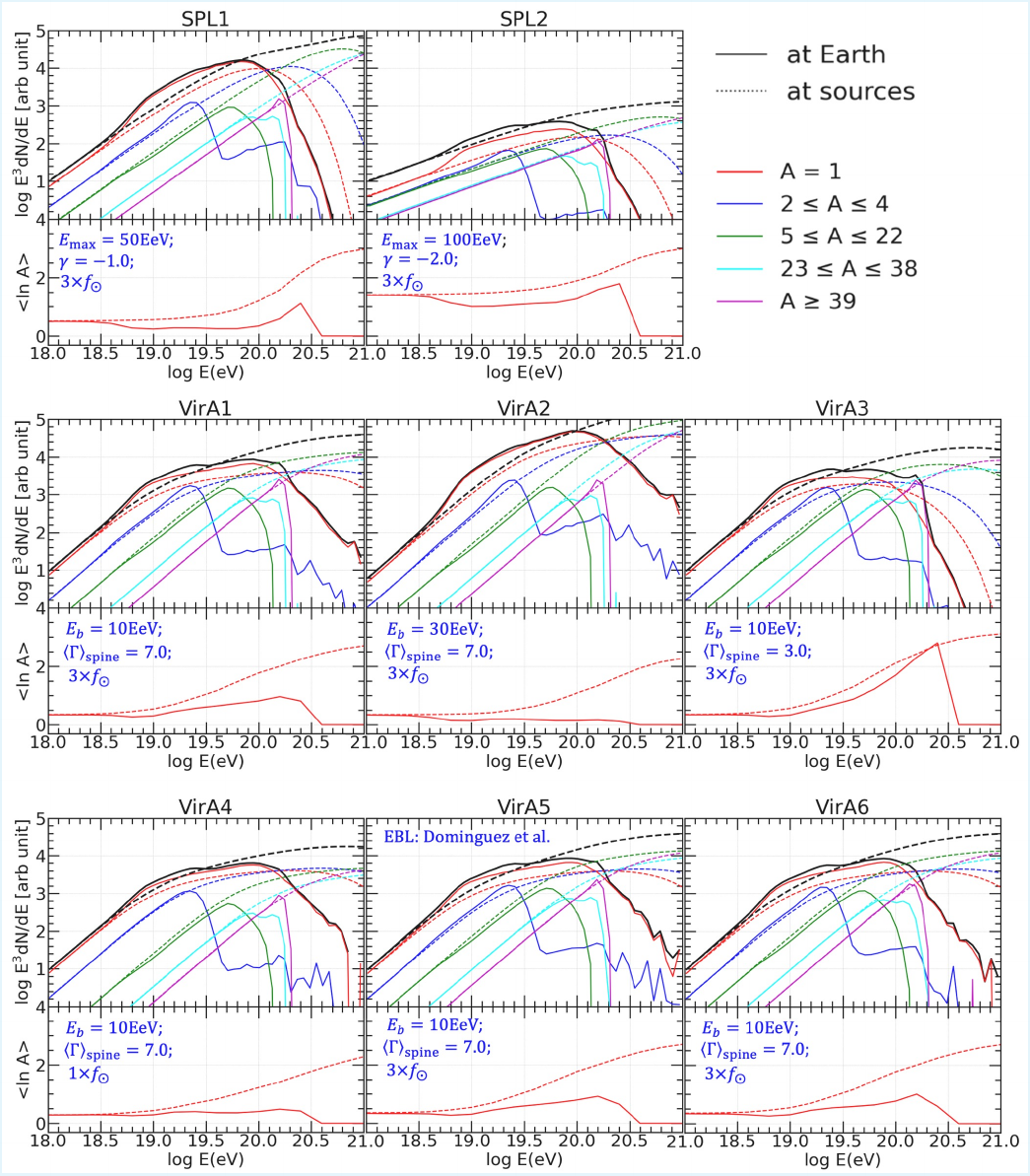}
\vskip -0.1 cm
\caption{Energy spectrum and mean logarithmic mass: $E_0^3dN/dE_0$ and $\langle \ln A_0 \rangle$ at the source (dashed lines) and $E^3dN/dE$ and $\langle \ln A\rangle$ at Earth (solid lines) after propagation for the two SPL models and the six Virgo A models in Table \ref{t2}. The relative amplitudes of the energy spectrum at sources are scaled such that their energy-integrated values are proportional to $Q_j/d^2$, except for the SPL models, for which the amplitudes at 1 EeV are normalized to match that of VirA1. In the energy spectrum plots, the colored lines represent the energy spectra for particles within different mass groups, while the black lines correspond to the total spectra. In the mean logarithmic mass plots, due to the small number of survived heavy nuclei, the values of $\langle \ln A \rangle$ at Earth (red solid lines) are reliably calculated only up to $E \sim 10^{20.3}$ eV and are artificially flattened above this energy.} 
\label{fig4}
\vskip 0.1 cm
\end{figure*}

\begin{figure*}[t]
\vskip 0 cm
\hskip 0 cm
\centering
\includegraphics[width=1\linewidth]{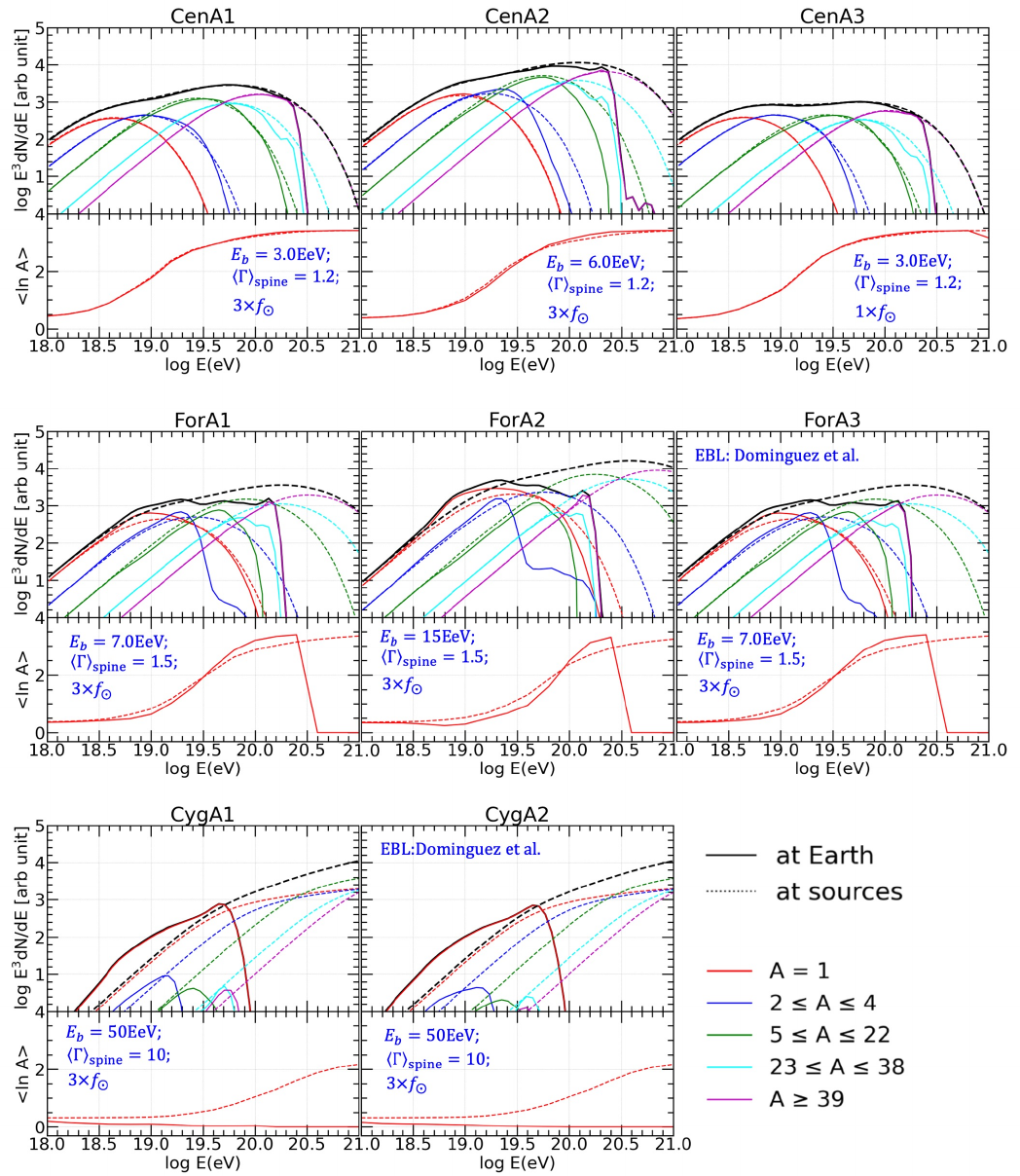}
\vskip -0.1 cm
\caption{Same as Figure \ref{fig4} except that the Centaurus A, Fornax A, and Cygnus A jet models in Table \ref{t2} are shown.}
\label{fig5}
\vskip 0.1 cm
\end{figure*}

\begin{figure*}[t]
\vskip -0.2 cm
\hskip 0 cm
\centering
\includegraphics[width=1\linewidth]{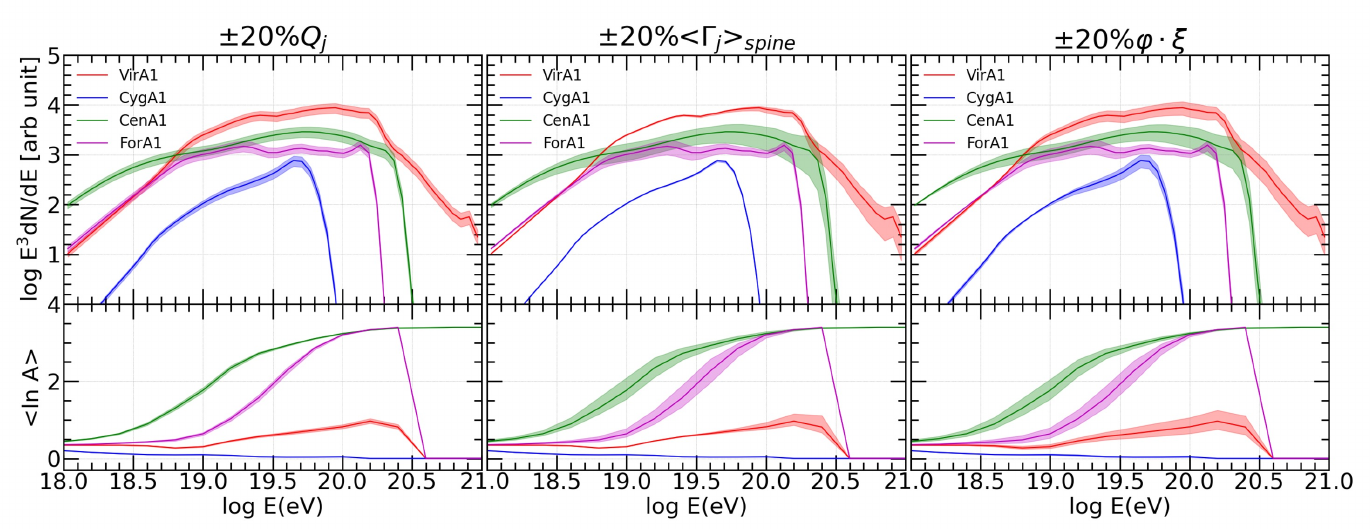}
\vskip -0.1 cm
\caption{Energy spectrum, $E^3dN/dE$, and mean logarithmic mass, $\langle \ln A \rangle$, of UHECRs reaching Earth from the four prominent RGs: Virgo A (red), Centaurus A (green), Fornax A (purple), and Cygnus A (blue). The solid lines show these quantities for the representative jet models, VirA1, CenA1, ForA1, and CygA1 in Table \ref{t2}. The shaded regions cover the ranges of changes when $Q_j$ (left), $\langle\Gamma\rangle_{\rm spine}$ (middle), and $\varphi \cdot \xi$ (right) are varied by $\pm 20\%$.}
\label{fig6}
\vskip 0.1 cm
\end{figure*}

\section{Energy Spectrum and Mass Composition of UHECRs at Earth}\label{s3}

We present the analyses of the energy spectrum and mass composition of UHECRs reaching Earth in CRPropa simulations. Our investigation focuses on the following: (1) the impact of our new DPL spectrum on the modification of these quantities due to propagation over the intergalactic and Galactic space, in comparison with the conventional SPL spectrum; (2) the effects of variations in different model parameters in the DPL spectrum; and (3) the comparison of the energy spectra and mass compositions of UHECRs from prominent RGs listed in Table \ref{t1}.

To illustrate the modification of the energy spectrum, we show
\begin{equation}
\begin{aligned}
\frac{dN}{dE}(E,A,Z) = \frac{dN}{dE_0}(E_0,A_0,Z_0) \times \\
\langle P(E_0,A_0,Z_0,E,A,Z,d_p) \rangle,
\end{aligned}
\end{equation}
referred to as the energy spectrum at Earth. Here, $dN/dE_0(E_0,A_0,Z_0)$ denotes the injection energy spectrum at the source, which is a simplified form of the expressions of Equations (\ref{DPLexp}) and (\ref{SPL}). Below, the energy spectrum is presented in arbitrary units. However, the relative amplitudes of $dN/dE_0(E_0,A_0,Z_0)$ for the jet models in Table \ref{t2} are scaled such that their energy-integrated values are proportional to $Q_j/d^2$, except for the SPL models, for which the amplitude at 1 EeV is aligned with that of VirA1, as described above. This allows for a direct comparison of the relative flux of UHECRs arriving at Earth from different RGs. The mean logarithmic masses, $\langle \ln A_0 \rangle$ at the source and $\langle \ln A \rangle$ at Earth, are calculated using $dN/dE_0(E_0,A_0,Z_0)$ and $dN/dE(E,A,Z)$, respectively.

Figure \ref{fig4} presents $dN/dE_0$ and $\langle \ln A_0 \rangle$ at the source (dashed lines) and $dN/dE$ and $\langle \ln A \rangle$ at Earth (solid lines) for the two SPL models and six VirA models listed in Table \ref{t2}. We note that in the $\langle \ln A \rangle$ plots, the values at Earth (red solid lines) are reliably calculated only up to $E \sim 10^{20.3}$ eV, due to the limited number of surviving heavy nuclei at such high energies. Compared to both SPL1 and SPL2, the VirA1 model features extended high-energy tails for all nucleus species at the source. As a result, the propagated spectrum $dN/dE$ for VirA1 (black solid line) exhibits a significantly enhanced high-energy tail above $E \sim 10^{20.2}$ eV, relative to the SPL models.

Focusing on VirA1, the representative model of Virgo A, we first point out the following general trends for the modification of the energy spectrum and mass composition of UHECRs.\hfil\break
(1) Energy spectrum modification: Due to energy losses via interactions with the CMB and EBL, as well as the photo-disintegration of heavy nuclei, the high-energy portions of $dN/dE$ decrease for all nucleus species, with heavier nuclei experiencing more severe reductions. Consequently, the total $dN/dE$ (black solid line) drops at $E\sim10^{20.2}$ eV, accompanied by an enhancement at lower energies, relative to the injection spectrum (black dashed line). However, in our DPL model for VirA1, characterized by an “extended” exponential cutoff in the source spectrum, some proton fragments from the disintegration of heavy nuclei retain energies above $E \gtrsim 10^{20.2}$ eV. Owing to the relatively short source distance of $d = 16$ Mpc, this results in a noticeable high-energy tail in the propagated proton spectrum (red solid line) beyond $E \sim 10^{20.2}$ eV.\hfil\break
(2) Mass composition modification: The photo-disintegration of nuclei during propagation converts heavier species into lighter ones with lower energies. Combined with energy losses, this alters the relative abundances of nuclei in a manner that is both complicated and energy dependent. As a result, the modification of $\langle \ln A \rangle$ can be non-trivial. In the case of the VirA1 model, heavy nuclei are significantly reduced at high energies ($E \gtrsim 10^{19.3}$ eV), while the proton fraction is enhanced across a wide energy range ($E \sim 10^{18.5} - 10^{20.2}$ eV). Hence, while the high-energy tail ($E \gtrsim 10^{20.2}$ eV) of the injection spectrum is dominated by heavy elements, the propagated spectrum at Earth in this energy range is primarily composed of protons. Correspondingly, $\langle \ln A \rangle$ decreases for $E \gtrsim 10^{18.5}$ eV. In general, $\langle \ln A \rangle$ tends to decrease due to the disintegration of nuclei during propagation, though it may also increase depending on the propagation distance and the initial source spectrum (see below).

The remaining VirA models in Figure \ref{fig4} illustrate the effects of varying key model parameters.\hfil\break
(1) Break energy: A comparison between VirA1 ($E_b = 10$ EeV) and VirA2 ($E_b = 30$ EeV) demonstrates that increasing $E_b$ shifts the exponential cutoff in $dN/dE_0$ to higher energies. Concurrently, $\langle \ln A_0 \rangle$ decreases above $E \sim 10^{18.8}$ eV, reflecting a lighter composition at the source. These changes result in a higher $dN/dE$ and a lower $\langle \ln A \rangle$ at Earth across the corresponding energy range.\hfil\break
(2) Jet Lorentz factor: Comparing VirA1 ($\langle\Gamma\rangle_{\rm{spine}} = 7$) with VirA3 ($\langle\Gamma\rangle_{\rm{spine}} = 3$) shows that a lower $\langle\Gamma\rangle_{\rm{spine}}$ shifts the exponential cutoff in $dN/dE_0$ to lower energies. Simultaneously, $\langle \ln A_0 \rangle$ increases above $E \sim 10^{19.3}$ eV, reflecting a heavier source composition at high energies. These differences have pronounced effects on the UHECRs reaching Earth: in the VirA3 model, $dN/dE$ declines steeply above $E \sim 10^{20.2}$ eV, and $\langle \ln A \rangle$ is significantly larger above $E \sim 10^{19.5}$ eV. The contrasting behaviors of VirA2 and VirA3 highlight the strong sensitivity of $dN/dE$ and $\langle \ln A \rangle$ at Earth to the modeling of the source spectrum, especially the role of an “extended” exponential cutoff.\hfil\break
(3) Metallicity: The comparison between VirA1 ($3\times f_{\odot}$) and VirA4 ($1\times f_{\odot}$) reveals that lower metallicity leads to a reduced $\langle \ln A_0 \rangle$ at the source, which in turn results in lower $\langle \ln A \rangle$ at Earth for $E \gtrsim 10^{19.2}$ eV.\hfil\break
(4) EBL model: The comparison between VirA1 and VirA5 indicates that differences in the adopted EBL models have only a marginal impact, primarily because Virgo A is a relatively nearby source.\hfil\break
(5) Propagation distance range: Comparing VirA1 and VirA6 shows that doubling the propagation distance range leads to a further suppression of $dN/dE$ at Earth, especially at the high-energy end. However, the effect on $\langle \ln A \rangle$ remains relatively minor.

Figure \ref{fig5} presents the same quantities as Figure \ref{fig4}, but for the CenA, ForA, and CygA models. It further illustrates the effects of varying model parameters.\hfil\break
(1) CenA and ForA models with $\langle\Gamma\rangle_{\rm{spine}}=1.2-1.5$: The effects of the extended exponential cutoff at $Z_0E_b \langle\Gamma\rangle_{\rm{spine}}^2$ are much less pronounced than in the VirA models with $\langle\Gamma\rangle_{\rm{spine}}=7$. As a result, $\langle \ln A \rangle$ tends to be heavier for these models.\hfil\break
(2) CenA models: Owing to the short distance of $d=4$ Mpc, the modifications of the energy spectrum and composition are relatively minor. However, heavy species belonging to the group of $A\geq39$ (purple lines), which dominate the high-energy end of the source spectrum, undergo severe losses for $E \gtrsim 10^{20.4}$, leading to a steep drop in the propagated spectrum $dN/dE$ at Earth above this energy. The comparison between CenA1 and CenA2 shows that increasing $E_b$ shifts the exponential cutoff in the source spectrum to higher energies, thereby extending $dN/dE$ at Earth and reducing $\langle \ln A \rangle$ in the range $E \sim 10^{18.3}$–$10^{20}$~eV. Similarly, the comparison between CenA1 and CenA3 indicates that lowering the metallicity reduces $dN/dE$ for $E\gtrsim 10^{19}$~eV and decreases $\langle \ln A \rangle$ in the range $E \sim 10^{18.5}$–$10^{19.5}$ eV, with negligible effects outside this range. The comparisons among these CenA models underscore that $\langle \ln A \rangle$ depends not only on the metallicity, but also strongly on the shape of the injection spectrum. Variations in $E_b$ and $\langle\Gamma\rangle_{\rm{spine}}$ within the DPL framework can lead to significant changes in both $dN/dE$ and $\langle \ln A \rangle$ at Earth.\hfil\break
(3) ForA models: Due to the relatively large distance of $d = 25$ Mpc, the total $dN/dE$ at Earth declines steeply above $E \sim 10^{20.2}$ eV. In addition, $\langle \ln A \rangle$ becomes slightly heavier in the energy range $E \sim 10^{19.5}$–$10^{20.5}$ eV, primarily as a result of photo-disintegration of heavy nuclei during propagation. These spectral and compositional features differ notably from those of the VirA1 model, but closely resemble the behavior of VirA3, which adopts a lower jet Lorentz factor of $\langle\Gamma\rangle_{\rm{spine}} = 3$. Such comparisons further emphasize the critical role of the extended exponential cutoff in shaping the propagated UHECR spectrum within the DPL model framework. The comparison between ForA1 and ForA2 shows that increasing $E_b$ enhances $dN/dE$ over the range $E \sim 10^{18.7}$–$10^{20}$ eV, while simultaneously reducing $\langle \ln A \rangle$ in the intermediate energy range. Meanwhile, the comparison between ForA1 and ForA3 indicates that differences in the EBL models have only a minor effect on the ForA predictions, consistent with the source’s moderate distance.\hfil\break
(4) CygA models: These models feature the largest $Q_j$, $E_b$, and $\langle\Gamma\rangle_{\rm{spine}}$ among all cases considered. However, due to the large distance of $d = 250$ Mpc, UHECRs from this source experience significant energy losses and photo-disintegration. As a result, although $dN/dE_0$ extends to much higher energies, the propagated $dN/dE$ at Earth drops steeply above $E \sim 10^{19.7}$ eV. Moreover, the disintegration of heavy nuclei leads to a low fraction of heavy elements among UHECRs arriving at Earth, resulting in a small $\langle \ln A \rangle$. The comparison between CygA1 and CygA2 shows differences due to the choice of EBL model; unlike in the Virgo A and Fornax A cases, the EBL model has a noticeable effect here due to the greater propagation distance. Nevertheless, both the CygA cases yield $\langle \ln A \rangle \approx 0$ at Earth.

Figure \ref{fig6} compares $dN/dE$ and $\langle \ln A \rangle$ for the representative models of the four prominent RGs: VirA1, CygA1, CenA1, and ForA1. This comparison aims to assess the relative contributions of these RGs to the observed UHECRs at Earth. As a reminder, the amplitudes of the injection spectra, $dN/dE_0$, are scaled such that their energy-integrated values are proportional to $Q_j/d^2$. To illustrate the model’s parameter dependence, shaded regions around each line indicate the range of variation when the DPL model parameters for each RG are varied by $\pm 20\%$. The figure highlights several key points.\hfil\break
(1) In the Northern Hemisphere, Virgo A is expected to be the dominant source of UHECRs observed at Earth. Although Cygnus A is a powerful FR-II radio galaxy with the largest $Q_j$, $E_b$, and $\langle\Gamma\rangle_{\rm{spine}}$, its contribution is suppressed due to its large distance ($d=250$ Mpc). In the Southern Hemisphere, Centaurus A and Fornax A contribute comparably to observed UHECR flux. These results are consistent with previous studies \citep[see, e.g.,][]{eichmann2018,eichmann2022}.\hfil\break
(2) For Virgo A with a large $\langle\Gamma\rangle_{\rm{spine}}$, the DPL model predicts a noticeable high-energy tail in $dN/dE$ at Earth above $E \sim 10^{20.2}$ eV. In contrast, for Centaurus A and Fornax A, which have smaller values of $\langle\Gamma\rangle_{\rm{spine}}$, the flux at Earth drops sharply above $E \sim 10^{20.2}$–$10^{20.3}$ eV.\hfil\break
(3) Due to photo-disintegration during propagation, the UHECR composition from Virgo A is significantly altered by the time the particles reach Earth. At high energies, the heavy-nuclei component is greatly diminished, while the proton fraction is enhanced across a wide energy range. As a result, $\langle \ln A \rangle$ decreases for $E \gtrsim 10^{18.5}$ eV. In contrast, for Centaurus A, which lies just 4 Mpc away, propagation effects are minimal, and the source’s heavy composition is largely preserved at Earth. Fornax A shows moderate composition changes, but the heavy nuclei still dominate among the UHECRs arriving at Earth due to the softer source spectrum associated with its small $\langle\Gamma\rangle_{\rm{spine}}$. Consequently, UHECRs from Virgo A have a lighter composition at Earth than those from Centaurus A and Fornax A.\hfil\break
(4) These qualitative trends remain robust even under variations of $\pm 20\%$ in the key model parameters, as indicated by the shaded bands in the figure.\hfil\break
(5) The figure suggests that the observed UHECR energy spectra, especially at the highest energies, and mass compositions may differ between the Northern and Southern Hemispheres, if RGs account for a significant portion of the UHECR flux. However, a more comprehensive and quantitative evaluation of the RG-origin model will require incorporating additional sources, including other RGs shown in Figure \ref{fig1} as well as those located beyond 300 Mpc.

\section{Summary}\label{s4}

In our previous studies, which combined relativistic hydrodynamic (RHD) simulations of jet flows with Monte Carlo simulations of CR acceleration and transport, we demonstrated that the time-asymptotic energy spectrum of UHECRs escaping from FR-type radio galaxies (RGs) can be well approximated by a double power-law (DPL) form with an ``extended'' exponential cutoff \citep{seo2023, seo2024}. This cutoff occurs at a characteristic energy scaling as $ZE_b \langle\Gamma\rangle_{\rm{spine}}^2$, where $ZE_b$ is the rigidity-dependent break energy for nuclei of charge $Z$, and $\langle\Gamma\rangle_{\rm{spine}}$ is the mean Lorentz factor of the jet spine. In this paper, we employed this newly derived source spectrum to explore the RG origin of UHECRs, focusing on contributions from several prominent RGs: the nearby Virgo A, Centaurus A, and Fornax A, as well as the more distant and powerful Cygnus A.

We utilized the CRPropa code \citep{batista2016, batista2022} to simulate the propagation of UHECRs from the RGs to Earth, accounting for interactions in the intergalactic and Galactic space. Instead of modeling full 3D trajectories through intervening magnetic fields, we adopted a more computationally efficient approach by simulating a statistical ensemble of 1D propagation paths across a range of source distances. From these simulations, we estimated the propagated energy spectrum, $dN/dE$, and mass composition, $\langle \ln A \rangle$, of UHECRs arriving at Earth. Our analysis focused on three main aspects: (1) the impact of the newly introduced DPL source spectrum---featuring an extended exponential cutoff---on $dN/dE$ and $\langle \ln A \rangle$, in comparison with the conventional single power-law spectrum (SPL); (2) the effects of varying model parameters in the DPL spectrum; and (3) the relative contributions to $dN/dE$ from prominent RG sources, and the modifications of $\langle \ln A \rangle$.

Our main findings are summarized as follows:\hfil\break
1) During propagation, UHECRs lose energy through interactions with the cosmic microwave background (CMB) and the extragalactic background light (EBL), as well as via the photo-disintegration of heavy nuclei. These processes significantly suppress the high-energy tail of $dN/dE$ observed at Earth. However, the shape of $dN/dE$ depends not only on the source distance but also on the form of the source spectrum. For Virgo A, the DPL model with an extended exponential cutoff predicts a pronounced high-energy tail of $dN/dE$ above $E \sim 10^{20.2}$eV, a feature not present in the corresponding SPL models. In contrast, for sources such as Centaurus A and Fornax A, both characterized by lower $\langle\Gamma\rangle_{\rm{spine}}$, the DPL model predicts a sharp decline in $dN/dE$ above $E \sim 10^{20.2}$–$10^{20.3}$eV.\hfil\break
2) The photo-disintegration of nuclei converts heavier elements into lighter ones, typically with lower energies. Together with energy losses, this alters the mass composition of UHECRs in a non-trivial way, depending on both the source distance and the injection spectrum. For Virgo A, our DPL model predicts a substantial reduction in $\langle \ln A \rangle$ at Earth for $E \gtrsim 10^{18.5}$eV, reflecting the loss of heavy nuclei during propagation. For Fornax A, despite more significant disintegration due to its longer distance, $\langle \ln A \rangle$ at Earth remains relatively similar to that at the source. In the case of Centaurus A, located only 4 Mpc away, the composition is largely preserved, and $\langle \ln A \rangle$ at Earth closely follows the initially heavy source composition.\hfil\break
3) Cygnus A, despite being a powerful FR-II RG, contributes only weakly to the observed UHECRs. This is primarily due to its large distance ($d = 250$ Mpc), which leads to severe energy losses and extensive nuclear disintegration during propagation, resulting in a significantly diminished flux at Earth.\hfil\break
4) Although our DPL models rely on several parameters, which carry some degree of arbitrariness, our principal conclusions remain robust across a range of plausible parameter values.\hfil\break
5) To assess the sensitivity of our results to various assumptions in the CRPropa simulations, we explored several variations. First, we tested two metallicity scenarios for the ISM in RG hosts: our default model assumes a heavy-element enhancement by a factor of three over the solar metallicity, while an alternative assumes the solar value. As expected, the lower metallicity yields a systematically smaller $\langle \ln A \rangle$ across all energies. However, $\langle \ln A \rangle$ depends not only on the source metallicity, but also strongly on the shape of the injection spectrum as well as the source distance, as mentioned above. Second, we examined the influence of different EBL models, comparing those of \citet{gilmore2012} and \citet{dominguez2011}, and found only marginal differences in the results. Finally, accounting for stronger magnetic deflections, up to twice the baseline level, leads to only minor changes in the predicted $dN/dE$ and $\langle \ln A \rangle$.

Our findings suggest that if RGs in the local universe are indeed significant sources of UHECRs, then the energy spectra and mass compositions of observed UHECRs may differ between the Northern and Southern Hemispheres, particularly at the highest energies. In the Northern Hemisphere, where Virgo A is expected to be the dominant source, UHECRs can exhibit a higher flux and a lighter mass composition at the high-energy end of the UHECR spectrum than in the Southern Hemisphere, where Centaurus A and Fornax A are the primary contributors. This hemispheric asymmetry arises in our DPL model due to differences in jet dynamics: Virgo A has a substantially more relativistic jet ($\langle\Gamma\rangle_{\rm{spine}} \approx 7$) than Centaurus A ($\langle\Gamma\rangle_{\rm{spine}} \approx 1.2$) and Fornax A ($\langle\Gamma\rangle_{\rm{spine}} \approx 1.5$).

However, to perform a comprehensive and quantitative assessment of the RG-origin scenario, it is necessary to include contributions from additional RGs, both those listed in Figure \ref{fig1} and others located beyond 300 Mpc. In any case, our findings underscore the importance of accurately characterizing the jet properties of nearby RGs and advancing our understanding of UHECR acceleration mechanisms in these sources. These efforts are crucial for evaluating the viability of RGs as dominant contributors to the observed UHECR population.

Finally, any search for the origins of UHECRs must be complemented by analyses of their arrival directions, particularly through studies of large-scale anisotropies \citep[see, e.g.,][]{PAO2007, abbasi2014, PAO2017}. Previous studies have explored whether observed anisotropies are consistent with the RG-origin scenario \citep[e.g.,][]{TA2022icrc, eichmann2022}. While we expect the DPL source spectrum introduced in this study to have only a modest impact on such correlation analyses---relative to the traditional SPL model---this hypothesis warrants a detailed investigation. We leave this important aspect for future work.

\begin{acknowledgments}
The authors would like to thank the anonymous referee for constructive comments and suggestions. JS’s work was supported by the Los Alamos National Laboratory (LANL) through its Center for Space and Earth Science (CSES). CSES is funded by LANL’s Laboratory Directed Research and Development (LDRD)program under project number 20240477CR. DR's work was supported by the National Research Foundation (NRF) of Korea through grant RS-2025-00556637, and HK's work was supported by NRF through grant RS-2023-NR076397.
\end{acknowledgments}

\bibliography{RadioGalaxy}{}
\bibliographystyle{aasjournal}

\end{document}